\documentclass[aps,prb,twocolumn,preprintnumbers,amsmath,amssymb,superscriptaddress]{revtex4-1}
\usepackage{graphicx}
\usepackage{subfigure}
\usepackage{epsfig}
\usepackage{dcolumn}
\usepackage{bm}
\usepackage{ulem}
\usepackage{color}
\usepackage{multirow}
\usepackage{slashed}

\def\avg#1{\left\langle#1\right\rangle}

\def\sgn{{\rm sgn}}

\def\be{\begin{equation}}       \def\ee{\end{equation}}
\def\bea{\begin{eqnarray}}      \def\eea{\end{eqnarray}}
\def\ba{\begin{array}}
\def\ea{\end{array}}
\def\bnum{\begin{enumerate} }
\def\enum{\end{enumerate}}

\def\nn{\nonumber}

\def\=>{\Rightarrow}
\def\>{\rightarrow}

\def\eye2{Fathbb{I}}

\renewcommand{\>}{\rangle}

\begin{document}

\title{Quantum criticality and duality in the Sachdev-Ye-Kitaev/AdS$_2$ chain}

\author{Shao-Kai Jian}
\email{jsk14@mails.tsinghua.edu.cn}
\affiliation{Institute for Advanced Study, Tsinghua University, Beijing 100084, China}
\affiliation{State Key Laboratory of Low Dimensional Quantum Physics, Tsinghua University, Beijing 100084, China}
\author{Zhuo-Yu Xian}
\email{xianzy@ihep.ac.cn}
\affiliation{Institute of High Energy Physics, Chinese Academy of Sciences, Beijing 100049, China}
\affiliation{School of Physics, University of Chinese Academy of Sciences, Beijing 100049, China}
\author{Hong Yao}
\email{yaohong@tsinghua.edu.cn}
\affiliation{Institute for Advanced Study, Tsinghua University, Beijing 100084, China}
\affiliation{State Key Laboratory of Low Dimensional Quantum Physics, Tsinghua University, Beijing 100084, China}

\begin{abstract}
We show that the quantum critical point (QCP) between a diffusive metal and ferromagnetic (or antiferromagnetic) phases in the SYK chain has a gravitational description corresponding to the double-trace deformation in an AdS$_2$ chain. Specifically, by studying a double-trace deformation of a $Z_2$ scalar in an AdS$_2$ chain where the $Z_2$ scalar is dual to the order parameter in the SYK chain, we find that the susceptibility and renormalization group equation describing the QCP in the SYK chain can be {\it exactly} reproduced in the holographic model. Our results suggest that the infrared geometry in the gravity theory dual to the diffusive metal of the SYK chain is also an AdS$_2$ chain. We further show that the transition in SYK model captures universal information about double-trace deformation in generic black holes with near horizon AdS$_2$ spacetime.
\end{abstract}
\date{\today}
\maketitle

\section{Introduction} 
The Sachdev-Ye-Kitaev (SYK) model exhibits many  structures and properties similar to a black hole \cite{kitaevtalk2015, Sachdev:1992fk, Maldacena:2016hyu}. For example, the low-energy symmetry structure in the SYK model leads to a nonconformal contribution to four-point functions captured by a Schwarzian derivative. This is general in any holographic system with a near-extremal black hole with nearly AdS$_2$ \cite{Maldacena:2016upp, Jensen:2016pah, Engelsoy:2016xyb}. It is worth emphasizing that the nonconformal contribution gives rise to an enhancement of out-of-time order correlators \cite{larkin1969, kitaevtalk2014, Shenker:2013pqa} with a saturated Lyapunov exponent \cite{kitaevtalk2015,Maldacena:2015waa}. Other propagating modes in four-point functions leave traces of the matter sector \cite{Maldacena:2016hyu, Polchinski:2016xgd, Jevicki:2016bwu, Gross:2017hcz}, which contains an infinite tower of particles. Recently, a three-dimensional bulk interpretation \cite{Das:2017pif} explains these propagating modes in terms of bilocal scalars. However, the dual theories of SYK models generalized to higher dimensions \cite{generalization} remain much less understood.

\begin{figure}
	\subfigure[]{\label{gravity}
		\includegraphics[width=3.5cm]{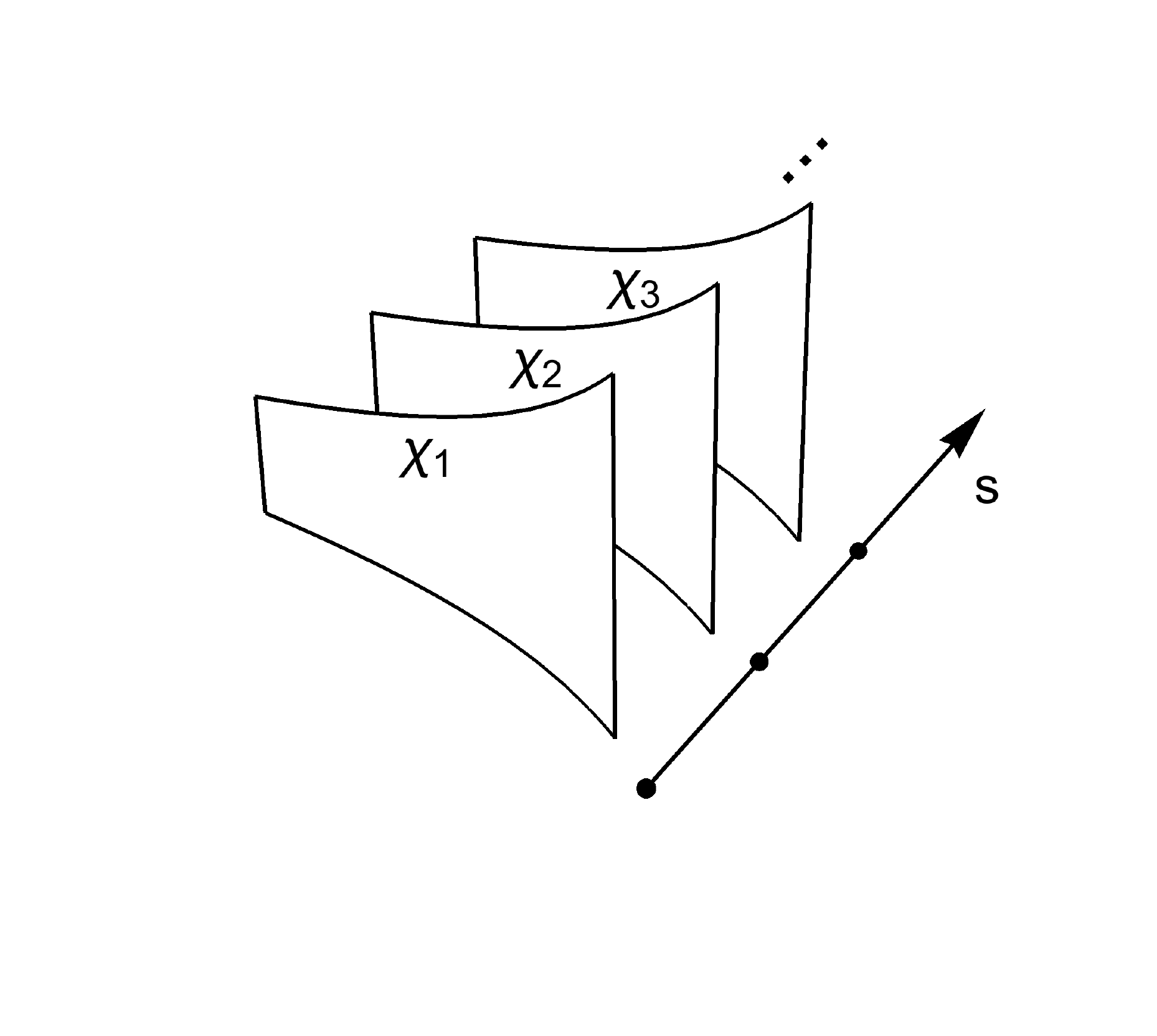}}~~~
	\subfigure[]{\label{phase_diagram}
		\includegraphics[width=3.5cm]{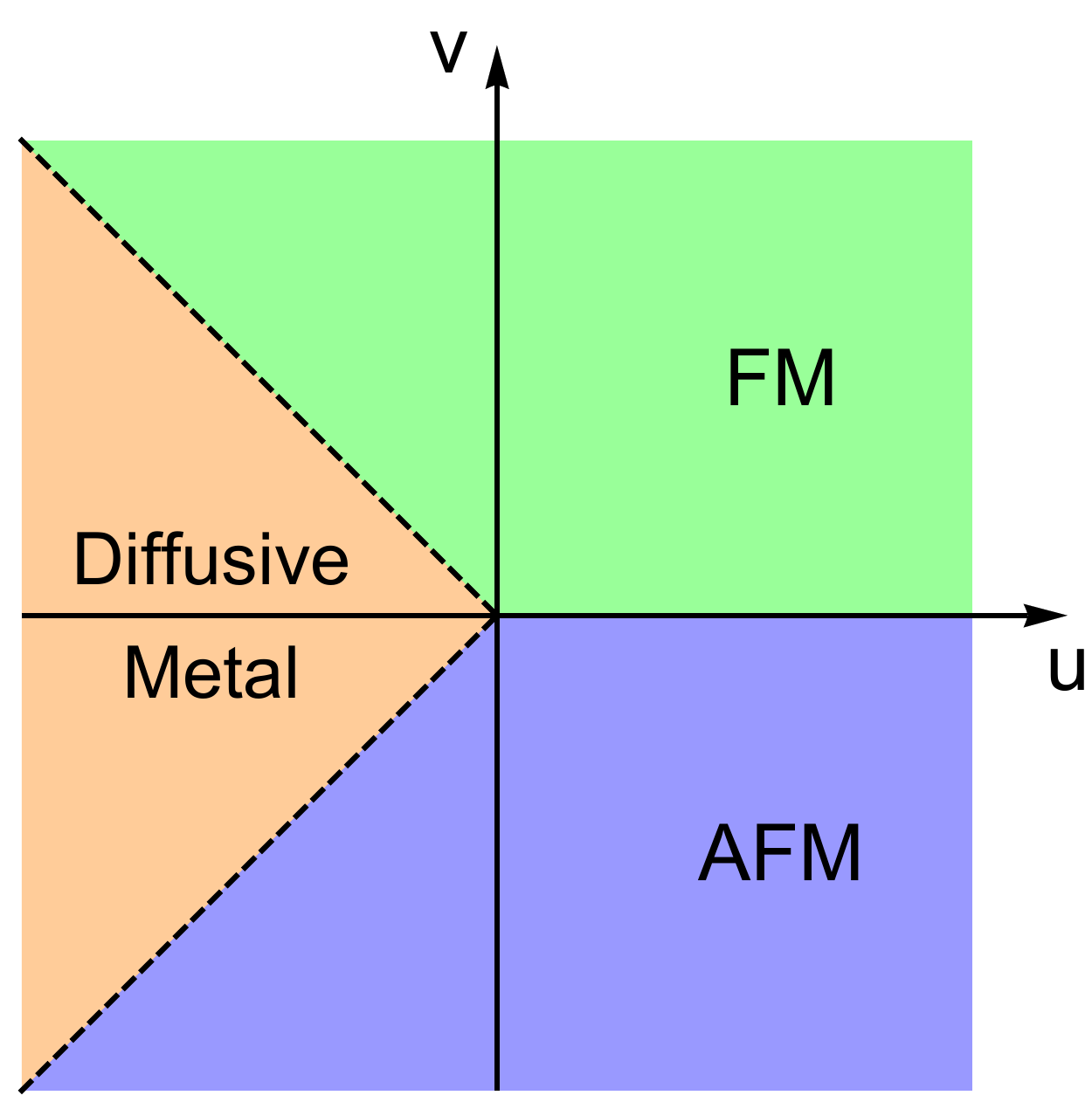}}
	\caption{(a) Schematic plot of the AdS$_2$ chain. $\chi_s$ denotes the scalar field dual to $O_s$ on site $s$ in the SYK chain. (b) The phase diagram of the generalized SYK chain with ``double-trace" deformations, and $q>4$. The dashed lines refer to phase boundaries between diffusive metal and FM/AFM. }
\end{figure}

In this paper, we show that the quantum critical point (QCP) in the SYK chain induced by certain type of interactions is dual to the holographic QCP in an AdS$_2$ chain induced by double-trace deformation. A schematic plot of an AdS$_2$ chain is shown in Fig. \ref{gravity}. While the diffusive metal in the SYK chain is time-reversal (TR) invariant, the ferromagnetic (FM)/antiferromagnetic (AFM) phase in the SYK chain breaks TR symmetry \cite{Bi:2017yvx, Jian:2017jfl}. We further show that the transition in the SYK model captures universal information about double-trace deformation in generic black holes with near horizon AdS$_2$ spacetime.

Double-trace deformation in AdS/CFT correspondence is a well known scenario to induce a nontrivial renormalization group flow from a CFT to either another CFT or to a symmetry breaking phase \cite{Witten:2001ua, Berkooz:2002ug, Sever:2002fk, Gubser:2002vv, Mueck:2002gm, Hartman:2006dy, Kaplan:2009kr, Iqbal:2011aj, Faulkner:2010gj, Faulkner:2010jy, Iqbal:2011ae}. These holographic QCPs rendered by double-trace deformation, sometimes called hybridized QCPs \cite{Iqbal:2011ae}, are argued to be dual to the spin-density wave (SDW) or nematic QCPs of metals associated with large Fermi surfaces and strong interactions \cite{sachdevbook, nematic, sdw, hertz1976, moriya1985, millis1993}. However, since there is no controllable calculation in the field theory side, the duality remains unclear.

Unlike the hybridized QCPs dual to SDW or nematic QCPs in metals, the calculations presented in this paper are under controlled on both sides of the duality, owing to the solvability of the SYK model. On the SYK side, we calculate the susceptibility near the QCP, which resembles the two-point function induced by double-trace deformation in gravity side. We also find that the nontrivial RG flow in the SYK chain triggered by the four-fermion interaction matches the double-trace flow in AdS/CFT correspondence. On the gravity side, we explicitly construct a holographic model, namely, a $Z_2$ scalar in the AdS$_2$ chain \cite{Kiritsis:2006hy, Fujita:2014mqa} with double-trace deformation at the AdS$_2$ boundary, to reproduce the same features of QCP in SYK chain, including the susceptibility and the RG equation. Our results strongly support that the low-energy geometry of the dual gravity theory of SYK chain should be taken as an AdS$_2$ chain.

\section{Generalized SYK chain and effective action} 
The generalized unperturbed SYK chain is given by the following Hamiltonian,
\bea
	H_0 &=& \sum_{s} \Big[ i^{\frac{q}{2}} \frac{J^0_{i_1 ... i_q,s}}{q!} \psi_{i_1}^s ...\psi_{i_{q}}^s \nn \\
	&&~~~~~~~~~~~~ + i^{\frac{q}{2}} \frac{J^1_{i_1 ... i_q,s}}{(\frac{q}{2}!)^2} \psi_{i_1}^s ...\psi_{i_{\frac{q}{2}}}^s \psi_{i_{\frac{q}{2}+1}}^{s+1} ...\psi_{i_{q}}^{s+1} \Big],
\eea
where $s$ is the site index. There are $N$ Majorana fermions in each site, denoted by $\psi_i^s$, $i=1,...,N$. $J^0_{i_1 ... i_q,s}$ and $J^1_{i_1 ... i_q,s}$ stand for the onsite and nearest-neighbor-site $q$-body Gaussian random interactions with mean zero and variances $\langle (J^0_{i_1 ... i_q,s})^2 \rangle= \frac{J_0^2(q-1)!}{N^{q-1}}$, $\langle (J^1_{i_1 ... i_q,s})^2 \rangle= \frac{J_1^2 (\frac{q}{2}!)^2 }{q N^{q-1}}$, respectively. The summation over $N$ fermion indices are implicit here and in the following. We assume $q=4 m$ where $m \ge1$ is an integer. It describes a diffusive metal with saturated Lyapunov exponent $\lambda= \frac{2\pi}{\beta}$ \cite{Gu:2016oyy}. The scaling dimension of Majorana fermion $\psi$ is given by $\Delta_f \equiv \frac1q$ in the IR. We will mainly focus on $q>4$, while the marginal case $q=4$ is given in the Appendix F. We consider four-fermion interactions (``double-trace" perturbations),
\bea
	H_2 &=& \sum_s \Big[ \frac{u}{8} C_{ij,s} C_{kl,s} \psi_i^s \psi_j^s \psi_k^s \psi_l^s \nn \\
	&&~~~~~~~~~~~~~~~~ + \frac{v}{8} C_{ij,s} C_{kl,s+1} \psi_i^s \psi_j^s \psi_k^{s+1} \psi_l^{s+1} \Big],
\eea
where $C_{ij,s}$ is a zero-mean Gaussian random variable with variance $\langle C_{ij,s}^2 \rangle =\frac{C^2}{N^2}$. $u$ and $v$ are tuning parameters in the system. 

We introduce a boson $O_s(\tau)$ to represent the order parameter, and a Lagrangian multiplier $B_s(\tau)$ to implement the identity $O_s=i \frac{C_{ij,s}}{2} \psi_i^s \psi_j^s$. Note that under TR transformation, $O_s$ field changes sign and serves as order parameter. We use replica trick to integrate out the random variables (see Appendix A). After introducing two bilocal bosonic fields $G^s$ and $\Sigma^s$, which are the propagator and self-energy of the Majorana fermions, and integrating out fermions, the replica averaged action is given by
\bea\label{averaged_action}
	S &=& \sum_s \Big[  \int d\tau (-\frac{u}{2} O_s^2-\frac{v}{2} O_s O_{s+1} + B_s O_s) \nn \\
	&& - \iint \frac{C^2}{4}  B_s(\tau_1)  G^s(\tau_1,\tau_2)^2 B_s(\tau_2) \Big] + S_0[G,\Sigma],
\eea
where $\iint \equiv \iint d\tau_1 d\tau_2$, and $S_0[G,\Sigma]$ is the effective action of the unperturbed SYK chain. Note that the replica index is omitted in large-$N$ limit \cite{Gu:2016oyy}. In the disordered phase, the equations of motion (EOM) is
\bea
	&& G^s(i\omega)^{-1} = -i\omega - \Sigma^s(i\omega),   \\
	&& \Sigma^s= J_0^2 (G^s)^{q-1}+ \frac{J_1^2}{2} (G^s)^{\frac{q}{2}}\big[ (G^{s-1})^{\frac{q}{2}-1}+ (G^{s+1})^{\frac{q}{2}-1} \big]. ~~~~
\eea
This recovers the EOM in the diffusive metal. In the conformal limit, the uniform saddle point solution is $G_c(\tau)=b \frac{\text{sgn}(\tau)}{|\tau|^{2\Delta_f}}$, with $b^q= \frac{(1-\frac2q)\tan\frac{\pi}{q}}{2\pi J^2} $, and $J^2\equiv J_0^2+J_1^2$.

\section{Susceptibility and RG equation}
The quadratic fluctuation of $B$ and $O$ fields around the saddle point solution will give rise to dynamic susceptibility $\mathcal{G}^R(t,p)\equiv i \theta(t) \langle [O_p(t), O_{-p}(0)] \rangle$ (see Appendix B). In frequency domain, it is given by
\bea\label{susSYK}
	\mathcal{G}^R(\omega,p)= \frac{1}{-u-v\cos p+ h(q) (-i\omega)^{1-4\Delta_f} },
\eea
where  $h(q)\equiv \frac{1}{ b^2C^2 \sin(2\pi \Delta_f) \Gamma(1-4\Delta_f) }>0$ and $p$ is the momentum. From the susceptibility (\ref{susSYK}), one can deduce that for $u<-|v|$, the diffusive metal is stable, while for $u>-|v|$, TR symmetry is spontaneously broken. (Saddle point equation analysis produces the results, see Appendix C.) For $v>0$, the susceptibility is $\mathcal{G}^R(0,p)^{-1}= -u-|v|+ \frac{|v|}{2}p^2$, and the ordered phase is FM; while for $v<0$, the susceptibility is $\mathcal{G}^R(0,p)^{-1}= -u-|v|+ \frac{|v|}{2}(p-\pi)^2$, and the ordered phase is AFM. The phase diagram is illustrated in Fig. \ref{phase_diagram}. Note that at the phase boundary between FM and AFM phases, i.e., $v=0$ in Fig. \ref{phase_diagram}, the averaged action (\ref{averaged_action}) has local $Z_2$ symmetry, $O_s \rightarrow \lambda_s O_s$, $\lambda_s=\pm 1$,  corresponding to $2^M$ degeneracies, where $M$ is the number of site.

Despite the difference in the ordering momentum, the FM and AFM are two similar phases since the disordered SYK chain explicitly breaks translation symmetry. Thus the transitions share the same universality class and the critical exponents are given by $[O]_\text{UV}=\frac2q$, $\nu_\text{crit}=\frac12$, $z_\text{crit}=\frac{2q}{q-4}$. Dictated by conformal symmetry at criticality, the finite temperature susceptibility is given by $\mathcal{G}^R(\omega,p, T)^{-1}= -u-|v|+ \frac{|v|}{2}\delta p^2 + T^{1-4\Delta_f}f(\frac{\omega}{T})$, where $f$ is a universal scaling function given in Appendix E and $\delta p$ is understood as small momentum fluctuations near the ordering momentum. One can see the critical temperature $T_c \sim |u+|v||^{\frac{q}{q-4}}$. The critical exponents of the $Z_2$ transition in the SYK chain are summarized in Appendix C.

The stability of the system against ``double-trace" perturbations, i.e.,  $-\sum_p \int \frac{d\omega}{2\pi}U_p |O_p|^2 $ (we have generalize the perturbation to a general function $U_p$) can also be captured by the following RG equation,
\bea\label{RGSYK}
	\frac{d U_p}{d\ln \mu}= -\Big( 1-\frac{4}{q}\Big)U_p - \frac{2}{h(q) \Lambda^{1-4\Delta_f}}U_p^2,
\eea
where $\Lambda$ is UV cutoff. Two fixed points include a critical one, $U_p=0$, corresponding to a phase transition and a stable one, $U_p=\frac{4-q}{2q}h(q) \Lambda^{4\Delta_f-1}$, corresponding to a diffusive metal. For $U_p=u+ v \cos p$, we can see that the phase boundary is at $u=-|v|$, consistent with the susceptibility. At $u=v=0$, one gets $[O_s]_\text{UV}=\frac2q$; the system has conformal symmetry in the low-energy limit. This fixed point is unstable upon the deformation, i.e., when $u<-|v|$, the system is driven to another fixed point with $[O_s]_\text{IR}=1-[O_s]_\text{UV}$, while the fermionic section remains largely unchanged and the system remains a diffusive metal. When $u>-|v|$ the symmetry is broken. This feature is similar to the double-trace deformation in AdS/CFT correspondence as we discuss below. It is worth emphasizing that the proper RG effect is captured by lowering energy with the chain length fixed, owing to the local criticality \cite{Gu:2016oyy} at low energy in an SYK chain. Then the scaling dimension of the Fourier component, denoted as $O_p$, is the same as $[O_s]$, and independent of momentum $p$. This important observation would lead to the conjecture that an AdS$_2$ chain is the proper IR geometry in the dual theory.

\section{Double-trace deformation in AdS/CFT duality} 
Inspired by the hints from the ``double-trace" perturbations in the SYK chains, we propose that $O_s$ can be viewed as a single-trace operator (it is {\it not} really a single-trace operator) that is dual to a scalar field $\chi_s$ in AdS$_2$, due to $O(N)$ freedoms in operator $O_s$. We further regconise the deformation $-\frac u2\int d\tau O_s^2$ as the double-trace deformation in AdS/CFT correspondence, which we briefly introduce now. Consider a $d$-dimensional Euclidean CFT$_{\text{UV}}$ containing a single trace operator $\tilde O$ whose scaling dimension is smaller than $d/2$, namely, $[\tilde O]_{\text{UV}}<\frac d2$. It corresponds to the undeformed SYK chain where $[O_s]<\frac12$ for $q>4$. The double-trace deformation, i.e., $-\frac{u}{2} \int d^dx \tilde O^2$, is relevant and will drive the system away.

When $u<0$, it ends up at CFT$_{\text{IR}}$ with $[\tilde O]_{\text{IR}}=d-[\tilde O]_{\text{UV}}$. The $\text{CFT}_\text{UV}$ and the CFT$_\text{IR}$ are related by a Legendre transformation at the large-$N$ limit \cite{Klebanov:1999tb}. One immediately recognizes that the two fixed points appeared in the RG equations (\ref{RGSYK}) of SYK chain correspond to CFT$_\text{UV/IR}$. Holographically, such deformation on the boundary CFT manifests itself as boundary conditions for the bulk field \cite{Witten:2001ua, Berkooz:2002ug, Sever:2002fk, Hartman:2006dy}. For the bulk field $\tilde\chi$ dual to $\tilde O$, the CFT$_\text{UV/IR}$ is dual to the bulk theory with alternative/standard quantization on $\tilde\chi$ \cite{Klebanov:1999tb}.

At large-$N$ limit, the deformed Green's function $\tilde{\mathcal{G}}_u(k)$ behaves as \cite{Iqbal:2011aj,Gubser:2002vv,Hartman:2006dy}
\bea\label{GuG0}
\tilde{\mathcal{G}}_u(k)=\frac1{-u+\tilde{\mathcal{G}}_0(k)^{-1}} \approx -\frac1u-\frac1{u^2 \tilde{\mathcal{G}}_0(k)},~~ |u \tilde{\mathcal{G}}_0(k)|\gg 1,~~~
\eea
where $k$ is the momentum. The susceptibility (\ref{susSYK}) in SYK chain is the non-local deformation version (nonvanishing $v$) of (\ref{GuG0}), whose derivation only relies on large-$N$ expansion \cite{Gubser:2002vv}, as is shown in Appendix D. The expansion of (\ref{GuG0}) at large $u$ tells the relation, $[\tilde O]_\text{UV}+[\tilde O]_\text{IR}=d$, which is also satisfied by the scaling dimensions of $O_s$ in the SYK chain in the $u=v=0$ UV fixed point and the IR fixed point in a diffusive metal.

When $u>0$, the double-trace deformation can stimulate instabilities in the bulk \cite{Iqbal:2011aj,Faulkner:2010gj} to symmetry breaking phases, which is consistent with the instability found in the SYK chain. However, the ordered phase in SYK chain is shown to be a nonchaotic thermal insulator with zero ground-state entropy and vanishing diffusion \cite{Jian:2017jfl}. It would be interesting to find a holographic counterpart of such thermal insulator and we leave it to future work.

\section{SYK/AdS$_2$ chain duality}
Here, we argue that it is more appropriate to take the low-energy geometry in the dual gravity theory of SYK chain as an AdS$_2$ chain than AdS$_2\times R$. The AdS$_2$ chain is a discrete set of AdS$_2$ spacetimes whose element is labeled by AdS$_{2,s}$, as shown in Fig. \ref{gravity}. When $u=v=J_1=0$, there is no interaction between different AdS$_{2,s}$'s and each AdS$_{2,s}$ is dual to the IR of a single SYK model at site $s$. When $J_1$ is turned on, AdS$_{2,s}$ develops interactions with AdS$_{2,s\pm1}$. We require that such interaction should not break the $SL(2,R)$ symmetry at each site. On the other hand, the duality between an SYK chain and the AdS$_2\times R$ is investigated in Ref. \cite{Davison:2016ngz}, where the space direction of the SYK chain is represented as the space $R$ in the bulk.

There are two evidences showing that the AdS$_2$ chain would be more promising than AdS$_2\times R$ as the IR geometry of an SYK chain. First evidence comes from symmetry consideration. The symmetry of the uniform saddle-point solution of the unperturbed SYK chain \cite{Gu:2016oyy} is $SL(2,R)^M\times Z_M$. It is the same as the isometry of a uniform AdS$_2$ chain, but different from the isometry $SL(2,R)\times R_T$ of AdS$_2\times R$, where the $R_T$ refers to spatial translation symmetry. (If the spatial-dependent axion field is considered, the spatial translation symmetry is also broken.) The second evidence comes from the scaling dimension of operator $O$. In the SYK chain, $[O_p]$ is independent of momentum $p$. The scaling dimension of an operator depends on the mass of its dual field in the bulk in holography. In the AdS$_2$ chain, the field dual to $O_p$ is the spatial Fourier transformation of $\chi_s$, namely, $\chi_p$. As we will see, the mass of $\chi_p$ is independent of $p$. However, in the AdS$_2\times R$, $O$ cannot be dual to a scalar field $\varphi$, since the Kaluza-Klein reduction of $\varphi$ on space $R$ gives a tower of $\varphi_p$ in the AdS$_2$ with momentum $p$ dependent mass \cite{Iqbal:2011aj,Iqbal:2011ae, Faulkner:2009wj}. The $p$ independence of $[O_p]$ causes the vanishing of correlation length in (\ref{susSYK}) when $v=0$, while the correlation length in AdS$_2\times R$ is finite.

\section{AdS$_2$ chain with double-trace deformation}
We consider a free $Z_2$ scalar field $\chi_s$ in the AdS$_{2,s}$ which is dual to the operator $O_s$ up to a normalized factor that would be determined later, i.e., $O_s=\zeta O_{\chi_s}$, where $O_{\chi_s}$ is the operator dual to $\chi_s$ in alternative quantization. We will study the linear perturbation of $\chi_s$ on a classical AdS$_2$ background and extract the Green's function of $O_s$ that gives the susceptibility.

The low-energy action for the AdS$_2$ chain is $S=S_{g,\phi}+S_\chi$, where $S_{g,\phi}=\sum_s (S_0[g_s,\phi_s]+S_I[g_s,\phi_s,g_{s+1},\phi_{s+1}])$. $g_s$ and $\phi_s$ refer to the metric and the dilaton on site $s$, respectively. $S_0[g_s,\phi_s]$ is Jackiw-Teitelboim action of dilaton-gravity theory \cite{Maldacena:2016upp, Teitelboim:1983ux, Jackiw:1984je}. While $S_I$ describes the inter-site coupling corresponding to a nonzero $J_1$ term in the SYK chain. $S_I$ should be properly designed, combining with $S_0[g_s,\phi_s]$, to give an AdS$_2$ chain ground state with the spatially uniform metric $ds^2_s=\frac{L^2}{z^2}(-dt^2+dz^2)$ and dilaton $\phi_s=\frac{\phi_r}z$, where $L$ is the AdS$_2$ radius and $\phi_r$ is the renormalized dilaton near the AdS$_2$ boundary.

While the action $S_{g,\phi}$ sets up the background geometry, the action for the scalars is
\bea
  S_\chi&=&\sum_s-\frac12\int dx^2\sqrt{-g_s} (g_s^{\mu\nu}\partial_\mu\chi_s\partial_\nu\chi_s + m^2 \chi_s^2 ),  \nn \\
	 &=&  \sum_p -\frac12\int d^2 x \sqrt{-g} \Big[ \partial_\mu \chi_{-p} \partial^\mu \chi_p + m^2 \chi_{-p} \chi_p   \Big],
\eea
where $\chi_p$ is the Fourier component of $\chi_s$ and the spatially uniform metric $g$ is used in the second line. Note that the mass of scalar is independent of momentum $p$. Let $\chi_p(t,z)=\chi_p(z)e^{-i\omega t}$, the EOM is
\bea
	\partial_z^2 \chi_p+ \Big(\omega^2- \frac{\nu^2- \frac14}{z^2} \Big) \chi_p=0, \label{EOM}
\eea
where $\nu= \sqrt{\frac14+m^2 L^2}$. Near the boundary, the asymptotic form of the scalar field is given by $\chi_p(z) \rightarrow A_p z^{\frac12- \nu} +B_p z^{\frac12+ \nu}$.

When $\nu \le 1$, alternative quantization can be applied. According to AdS/CFT duality, the generating functional $Z[J_s]$, where $J_s$ is the source of scalar $O_s$, is achieved by evaluating \cite{Faulkner:2010jy,Skenderis:2002wp}
\bea
  &\avg{e^{i\sum_s \int dt J_sO_s}}_{H_0+H_2}\approx \left.\lim\limits_{\epsilon\to0} e^{i(S_\chi+S_{ct}+S_{u,v}+S_J)}\right|_{\chi_{cl}},
\eea
with the boundary terms given by
\bea\label{actionscalar}
  S_{ct} &=&\frac{2\nu-1}4  \sum_s \int_{z=\epsilon} dt \sqrt{-h}\chi_s^2, \\
  S_{u,v} &=& \frac12\zeta^2\epsilon^{2\nu}\sum_s \int_{z=\epsilon} dt\sqrt{-h} (u\chi_s^2+v\chi_s\chi_{s+1}), \\
  S_J &=& \zeta\epsilon^{\frac12+\nu}\sum_s\int_{z=\epsilon} dt\sqrt{-h} J_s\chi_s,
\eea
where $h$ refers to the induced metric on the $z=\epsilon$ surface, and $\chi_{cl}$ refers to classical solution of $\chi$. The counter term $S_{ct}$ is introduced to cancel the divergence of $S_\chi$ on the boundary and $S_{u,v}$ corresponds to the double-trace deformation.

Requiring the in-going boundary condition near the horizon $z\to\infty$, we can extract the retarded Green's function of $O_p$ on the boundary \cite{Son:2002sd}, which is defined as $\mathcal{G}^R(t,p) \equiv i \theta(t) \langle [O_p(t), O_{-p}(0)] \rangle$. We firstly consider the case of $u=v=0$. The undeformed retarded Green's function is (see Appendix E) \bea
	\mathcal{G}^R_{0,0}(\omega,p) =-\frac{\zeta^2A_p(\omega)}{2\nu B_p(\omega)}=\zeta^2  \frac{2^{2\nu-1}\Gamma(\nu)}{\Gamma(1-\nu)}(-i\omega)^{-2\nu}.
\eea
It coincides with the undeformed susceptibility in the SYK chain provided $\nu=\frac12-\frac2q$ and $
\zeta^2=\frac{\sqrt{\pi } C^2 b^2 \Gamma \left(\nu +\frac{1}{2}\right)}{\tan (\pi  \nu )\Gamma (\nu )}\approx \frac{C^2}{\sqrt{4\pi J^2}} +O(\nu)$. Turning on the double-trace deformation, the Green's function becomes
\bea\label{Gu}
	\mathcal{G}^R_{u,v}(\omega,p) = \frac{1}{-u-v \cos(p)+ \mathcal{G}^R_{0,0}(\omega,p)^{-1}},
\eea
which matches (\ref{susSYK}) exactly! Thus, the (in)stability of the holographic model is the same as in the SYK chain. When $u<-|v|$, the IR is the AdS$_2$ chain with standard quantization. When $u>-|v|$, the scalar fields condense as $\chi_s=\sgn(v)\chi_{s+1}$, corresponding to the FM/AFM phase, respectively, and break the $Z_2$ symmetry. They will backreact to the background and lead to a new geometry in the IR, which waits for further study.

According to the AdS/CFT correspondence, a CFT at finite temperature is equivalent to the presence of a black hole in the bulk theory. Thus we consider an AdS$_2$ black hole background to find the finite temperature Green's function. Following the same steps, the Green's function at finite temperature $T$ is $\mathcal{G}^{R}_{0,0}(\omega,p,T) = T^{-2 \nu} f(\frac{\omega}{T})^{-1}$, where $f$ is the a universal scaling function given in Appendix E. And the deformed Green's function at finite temperature $\mathcal{G}^{R}_{u,v}(\omega,p, T)$ also satisfies (\ref{Gu}) with $\mathcal{G}^R_{0,0}(\omega,p)$ replaced by $\mathcal{G}^{R}_{0,0}(\omega,p,T)$. These formulas are exactly the same with the finite temperature susceptibility calculated in the SYK chain.

By exactly reproducing the susceptibility in SYK chain, we demonstrate the dual description of the QCP by a Z$_2$ scalar in an AdS$_2$ chain. Moreover, we also calculate the beta functions by the holographic RG method \cite{Faulkner:2010jy}, and for a general double-trace deformation, the result is
\bea
 \frac{d U_p}{d\ln \epsilon}&=&2\nu U_p + \epsilon_0^{2\nu}\zeta^2 U_p^2,
\eea
where $\epsilon_0^{-1}$ is the UV cutoff. With $\zeta^2 \propto h(p)^{-1}$, the RG equation in an SYK chain, (\ref{RGSYK}), is reproduced by the holographic method, which again strongly suggests the duality between them.

\section{Conclusion and discussion} In this paper, we study a QCP rendered by a particular kind of four-fermion interaction in the SYK chain and show that it has holographic description by double-trace deformation in an AdS$_2$ chain. Owing to the large-$N$ degrees of freedom on each site, such QCP is independent of the spatial dimension, and can be generalized to any dimensions. Our proposal also opens the door of experimental realization of double-trace deformation in gravity systems in the sense of AdS/CFT correspondence.

It is illuminating to compare the QCPs in the SYK chain and the nematic or SDW QCPs in metals. While both QCPs exhibit local quantum critical behavior, the origins are different: the local criticality comes from the large-$N$ degrees of freedom on each site in the SYK chain, on the other hand, the underlying Fermi surfaces are responsible for that of QCPs in metals. This difference also leads to different dual IR geometries. The dual IR geometries for nematic or SDW QCPs in metals are proposed to be AdS$_2\times R^2$ \cite{Faulkner:2010gj, Iqbal:2011aj, Iqbal:2011ae}.

The QCP in the SYK chain can also be understood by the semi-holographic effective theory \cite{Faulkner:2010tq,Jensen:2011af} consisting of a Landau-Ginzburg theory of an order parameter and an emergent conformal sector. The emergent large-$N$ degrees of freedom in the conformal sector can be identified as the $B$ field that interacts strongly with the Majorana fermions, as shown (\ref{averaged_action}). We represent the concrete semi-holographic effective theory in Appendix H.

We further consider double-trace deformations of a scalar field in a four-dimensional near extremal Reissner-Nordstrom AdS black hole (see Appendix G), and find that the Green's function is controlled by the nearly AdS$_2$ geometry in the vicinity of horizon. Since we have established the correspondence between the transition in the SYK model and in the AdS$_2$ spacetime, in this sense the transition in the SYK model captures the universal properties of double-trace deformations in generic near-extremal black holes with near horizon AdS$_2$ spacetime.

After establishing the duality between two-point functions, it is worth constructing the bulk interactions of scalar fields to match the multipoint function of the dual operators. Another extension is to consider a $2l$-fermion interaction $(C_{i_1,...,i_l}\psi_{i_1}...\psi_{i_l})^2$, where $l\le \frac{q}{2}$ is an even number. In such situation, there is also a transition to symmetry breaking phases; however, unlike in the case of four-fermion interaction, we conjecture that the ordered phase would still be a chaotic phase with a saturated Lyapunov exponent, and consequently,  dual to the AdS$_2$ chain. It is possible that AdS$_2$ chain can be reduced from a Majumdar-Papapetrou solution in higher dimension, where our nearest-neighbor interactions come from the coupling between their throats \cite{Maldacena:1998uz,Myers:1986rx,Klebanov:1999tb}.

\begin{center}
{\bf ACKNOWLEDGEMENT}
\end{center}
We would like to thank Yi Ling, Wei Song, Hongbao Zhang, Xiao-Ning Wu, Hao Ouyang and Yi-Kang Xiao for helpful discussions. SKJ and HY are supported in part by the NSFC under Grant No. 11474175 and by the MOST of China under Grant No. 2016YFA0301001. ZYX is supported by the NSFC under Grant No. 11575195.

\begin{widetext}

\renewcommand{\theequation}{A\arabic{equation}}
\setcounter{equation}{0}
\renewcommand{\thefigure}{A\arabic{figure}}
\setcounter{figure}{0}
\renewcommand{\thetable}{A\arabic{table}}
\setcounter{table}{0}

\begin{center}
{\bf APPENDIX}
\end{center}

\subsection{Effective action of SYK chain}
The Lagrangian is
\bea
	L &=& \sum_{s} \Big[ \frac12 \psi_i^s \partial_\tau \psi_i^s+  i^{\frac{q}{2}} \frac{J^0_{i_1 ... i_q,s}}{q!} \psi_{i_1}^s ...\psi_{i_{q}}^s + i^{\frac{q}{2}} \frac{J^1_{i_1 ... i_q,s}}{(\frac{q}{2}!)^2} \psi_{i_1}^s ...\psi_{i_{\frac{q}{2}}}^s \psi_{i_{\frac{q}{2}+1}}^{s+1} ...\psi_{i_{q}}^{s+1} \\
	&& + \frac{u}{8} C_{ij,s} C_{kl,s} \psi_i^s \psi_j^s \psi_k^s \psi_l^s + \frac{v}{8} C_{ij,s} C_{kl,s+1} \psi_i^s \psi_j^s \psi_k^{s+1} \psi_l^{s+1}\Big],
\eea
where the summation over the Majorana index is implicit. Introducing a boson $O_s(\tau)= \frac{i C_{ij,s}}{2} \psi_i^s(\tau) \psi_j^s(\tau)$ to decouple the double-trace term and a Lagrange multiplier $B_s$, we arrive at
\bea
	L &=& \sum_{s} \Big[  \frac12 \psi_i^s \partial_\tau \psi_i^s+ i^{\frac{q}{2}} \frac{J^0_{i_1 ... i_q,s}}{q!} \psi_{i_1}^s ...\psi_{i_{q}}^s + i^{\frac{q}{2}} \frac{J^1_{i_1 ... i_q,s}}{(\frac{q}{2}!)^2} \psi_{i_1}^s ...\psi_{i_{\frac{q}{2}}}^s \psi_{i_{\frac{q}{2}+1}}^{s+1} ...\psi_{i_{q}}^{s+1} - \frac{u}{2} O_s^2- \frac{v}{2} O_s O_{s+1} + B_s(O_s- i  \frac{C_{ij,s}}{2} \psi_i^s \psi_j^s)  \Big], \nn \\
\eea
By using the replica trick, we average over the disorder
\bea
	S &=& \sum_s \Big( \int d\tau \big[ \frac12 \psi_i^s \partial_\tau \psi_i^s - \frac{u}{2} O_s^2- \frac{v}{2} O_s O_{s+1} + B_s O_s \big] - \frac{C^2}{4N^2} \iint d\tau_1 d\tau_2  B_s(\tau_1) B_s(\tau_2) \big( \psi_i^s(\tau_1) \psi_i^s(\tau_2)\big)^2 \\
	&& -  \frac{1}{2q N^{q-1}}\iint d\tau_1 d\tau_2 \big[ J_0^2 \big(\psi_{i}^s(\tau_1)\psi_{i}^s(\tau_2)\big)^q+ J_1^2 \big(\psi_{i}^s(\tau_1)\psi_{i}^s(\tau_2)\big)^{\frac{q}{2}} \big(\psi_{j}^{s+1} (\tau_1)\psi_{j}^{s+1}(\tau_2)\big)^{\frac{q}{2}} \big] \Big).
\eea
Note that only the replica diagonal part survives in the large-$N$ limit. Introducing bi-local bosons $G^s(\tau_1,\tau_2)= \frac1N \psi_i^s(\tau_1) \psi_i^s(\tau_2)$ and self-energy $\Sigma^s(\tau_1,\tau_2)$ to decouple the interactions \cite{Gu:2016oyy}, and integrating out Majorana fermions, the action reads
\bea
	S &=& \sum_s \Big(  \int d\tau [-\frac{u}{2} O_s^2- \frac{v}{2} O_s O_{s+1}+ B_sO_s] - \iint d\tau_1 d\tau_2 \frac{C^2}{4}  B_s(\tau_1)  G^s(\tau_1,\tau_2)^2 B_s(\tau_2)\Big) + S_0[G,\Sigma],
\eea
where $S_0[G,\Sigma]$ is the effective action of the unperturbed SYK chain,
\bea
	S_0[G,\Sigma] &=& \sum_s \Big(- \frac{N}{2} \text{Tr} \ln(\partial_\tau-\Sigma^s)+ \frac{N}{2}\iint d\tau_1 d\tau_2 \Big[ \Sigma^s(\tau_1,\tau_2) G^s(\tau_1,\tau_2)- \frac1q \big[ J_0^2 G^s(\tau_1,\tau_2)^q+ J_1^2 G^s(\tau_1,\tau_2)^{\frac{q}{2}} G^{s+1} (\tau_1,\tau_2)^{\frac{q}{2}} \big] \Big] \Big). \nn
\eea
\subsection{The dynamic susceptibility in SYK chain}
The equations of motion (EOM) are
\bea
	&& uO_s(\tau)+ \frac{v}{2}[O_{s-1}(\tau)+O_{s+1}(\tau)]=B_s(\tau),~~~ O_s(\tau)= \frac{C^2}{2}\int d\tau' G^s(\tau',\tau)^2 B_s(\tau), ~~~ G^s(i\omega)^{-1} = -i\omega - \Sigma^s(i\omega),  \\
	&& \Sigma^s(\tau_1,\tau_2)= J_0^2 G^s(\tau_1,\tau_2)^{q-1}+ \frac{J_1^2}{2} G^s(\tau_1,\tau_2)^{\frac{q}{2}-1}\big[ G^{s-1}(\tau_1,\tau_2)^{\frac{q}{2}}+ G^{s+1}(\tau_1,\tau_2)^{\frac{q}{2}} \big]+ \frac{C^2}{N} G^s(\tau_1,\tau_2) B_s(\tau_1) B_s(\tau_2), \nn \\
\eea
In the disorder phase, we have $O_s=B_s=0$ and the EOM reduce to
\bea
G^s(i\omega)^{-1} = -i\omega - \Sigma^s(i\omega),~~~~ \Sigma^s= J_0^2 (G^s)^{q-1}+ \frac{J_1^2}{2} (G^s)^{\frac{q}{2}}\big[ (G^{s-1})^{\frac{q}{2}-1}+ (G^{s+1})^{\frac{q}{2}-1} \big].
\eea
In the conformal and uniform limit, we have
\bea
G_c(\tau) =b \frac{\text{sgn}(\tau)}{|\tau|^{2\Delta_f}},
\eea
where $b^q= \frac{(1-\frac2q)\tan\frac{\pi}{q}}{2\pi J^2} $, and $\Delta_f\equiv \frac1q$. In next step, we consider the fluctuations of $B$ and $O$ around this saddle point,
\bea
	S &=& \sum_s \Big(  \int d\tau (-\frac{u}{2} O_s^2- \frac{v}{2} O_s O_{s+1} + B_s O_s) - \iint \frac{C^2}{4}  B_s(\tau_1)  G_c(\tau_1,\tau_2)^2 B_s(\tau_2) \Big), \\
	 &=& \sum_p \int \frac{d\omega}{2\pi} \Big( O_{-p} (-\frac{u}{2} -\frac{v}{2} \cos p )O_p+ B_{-p} O_p - b^2 c(2\Delta_f) \frac{C^2}{4} B_{-p} (-\omega)  |\omega|^{4\Delta_f-1} B_p(\omega) \Big),
\eea
where $c(\Delta)=2 \sin(\pi \Delta) \Gamma(1-2\Delta)$ and $A_p= \frac{1}{\sqrt{M}} A_s e^{ips}$ is the Fourier component of field $A$, $M$ is the total site. One can integrate out $B$ fields, to get
\bea
	S &=& \frac 12 \sum_p \int \frac{d\omega}{2\pi} O_{-p}(-\omega) \Big[-u-v\cos p+ h(q) |\omega|^{1-4\Delta_f} \Big] O_p(\omega),
\eea
where $h(q)$ is given in the main text. After analytic continuation, the dynamic susceptibility is given by
\bea
	\mathcal{G}^R(\omega,p)= \frac{1}{-u-v\cos p+ h(q) (-i\omega)^{1-4\Delta_f} }.
\eea

\subsection{The transition at finite temperature in the SYK chain}

In the ordered phase, the order parameter $O$ develops long-ranged correlations \cite{Bi:2017yvx, Jian:2017jfl}, i.e., $\langle O_s(\tau_1)O_{s'}(\tau_2)\rangle=N \bar O^2$, where $\bar O$ is a constant (assuming $v>0$ here), then we have,
\bea
	&& G_f(i\omega)^{-1} = -i\omega - \Sigma_f(i\omega), \\
	&& \Sigma_f(\tau)= J^2 G_f(\tau)^{q-1}+ (u+v)^2 \bar O^2 C^2 G_f(\tau),  \\
	&& \bar O= \frac{(u+v)C^2}{2} \bar O \int d\tau' G_f(\tau')^2.
\eea
From the last equation, clearly, for $u<-|v|$, there is no solution. While for $u+v>0$, we get the ``gap equation", $1= \frac{(u+v)C^2}{2}\int d\tau' G_f(\tau')^2$. Here, we get two energy scales expressed $\Lambda_0=J^2/(u+v)C^2$ and $\Lambda_2=(u+v)\bar O^2$. In the small $\bar O$ limit, $\Lambda_0 \gg \Lambda _2$, among this large region, the propagator would behave like the conformal solution, thus
\bea
	1\approx \frac{(u+v)C^2}{2}\int_{\Lambda_0^{-1}}^{\Lambda_2^{-1}} d\tau \frac{b^2}{\tau^{\frac4q}},
\eea
from which one can get $\bar O  \propto  |u+v|^{\frac{2}{q-4}}$. The critical exponent is $\beta=\frac{2}{q-4}$.

According to conformal symmetry, the finite temperature susceptibility is given by
\bea
	\mathcal{G}^R(\omega, p, T)= \frac{1}{-u-|v|+ \frac{|v|}{2}\delta p^2 + T^{1-4\Delta_f}f(\frac{\omega}{T})},
\eea
where $f$ is a universal scaling function. It is given by $f^{-1}(\frac{\omega}{T})=- \frac{\zeta^2 T^{2\nu}}{2\nu} \eta(\omega, \frac{1}{2\pi T})$ in Appendix E. Then $\mathcal{G}^R(0,0,T)=\frac{1}{-u-|v|+f(0)T^{1-4\Delta_f}}$, giving rise to $T_c \sim |u+v|^{\frac{q}{q-4}}$. A summary of critical exponents of the $Z_2$ transition in the SYK chain is shown in Table A1.

\begin{table}\label{exponent}
  \centering
  \caption{The critical exponents for the $Z_2$ transitions in the SYK chain and the semi-holographic effective theory. $q$ refers to the $q$-body interaction in the SYK chain, and $\Delta$ refers to the scaling dimension of the emergent large-$N$ field in the conformal sector in the semi-holographic effective theory. The critical exponents are provided $\Delta=1-\frac2q$.}
    \begin{tabular}{c c c c c c c c c}
    \hline
    \hline
     Critical exponents & ~~ $\alpha$~~   &  ~~$\beta$~~  & ~~$\gamma$~~ & ~~$\delta$~~ &  ~~$\nu$~~ & ~~$\eta$~~ & ~~$z$~~ & ~~$T_c$~~\\
     \hline
     SYK chain & $\frac{q-8}{q-4}$ & $\frac2{q-4}$ & 1 & $\frac{q-2}2$ & $\frac12$ & 0 & $\frac{2q}{q-4}$ & $\frac{q}{q-4}$\\
     Semi-holographic theory~~ & $\frac{4\Delta-3}{2\Delta-1}$ & $ \frac{1-\Delta}{2\Delta-1}$ & 1 & $\frac\Delta{1-\Delta}$ & $\frac12$ & 0 & $\frac2{2\Delta-1}$ & $\frac{1}{2\Delta-1}$\\
\hline
\hline
    \end{tabular}%
  \label{tab1}%
\end{table}%

\subsection{Green's function with non-local double-trace deformation}
The formula of double-trace deformation in field theory is given in Ref. \cite{Gubser:2002vv}. We will apply it to the case of non-local double-trace deformation and link it to the SYK chain in the main text.  The Green's function under a non-local double-trace deformation can be calculated by evaluating the Euclidean partition function
\bea
  Z_{U}[J]  = \avg{ e^{\int_k( \frac12 O_{-k} U_k O_{k} + J_{-k} O_k )}}_0,
\eea
where $\avg{\cdots}_0$ denotes the expectation value under the undeformed CFT. We work in momentum space and $\int_k$ is the abbreviation of $\int\frac{d^d k}{(2\pi)^d}$. We apply Hubbard-Stratonovich transformation by introducing a field $B$,
\bea
  Z_{U}[J] = \sqrt{\det(U^{-1})} \int\mathcal{D}B e^{ -\frac12\int_k B_{-k}U_k^{-1}B_{k} } \avg{ e^{\int_k (B+J)_{-k}O_{k}} }_0.
\eea
At large $N$, we assume that higher point functions of $O$ are suppressed, which leads to
\be\label{largeNsuppress}
  \avg{e^{\int_k B_{-k}O_{k}}}_0 \approx e^{\frac12\int_k B_{-k} \mathcal{G}_0(k) B_k},
\ee
where $\mathcal{G}_0(k)=\avg{O_{k}O_{-k}}_0$ is the undeformed Green's function. We will see that the above equation is true in the SYK chain where $O_s(\tau)=i\frac{C_{ij,s}}{2}\psi_i^s(\tau)\psi_j^s(\tau)$. By applying (\ref{largeNsuppress}),
\bea
  Z_{U}[J] &\approx& \sqrt{\det(U^{-1})} \int\mathcal{D}B e^{ -\frac12\int_k \big[-B_{-k}U_k^{-1}B_{k} +\int_k (B+J)_{-k} \mathcal{G}_0(k) (B+J)_k \big] }    \\
  &=&   \sqrt{\det\Big(\frac1{1-U^{-1}\mathcal{G}_0}\Big)} e^{\frac12\int_k J_{-k}\mathcal{G}_U(k)J_{k}},
\eea
where the deformed Green's function is
\be
    \mathcal{G}_U(k)=\frac1{-U_k+ \mathcal{G}_0(k)^{-1}},
\ee
which is the same as (\ref{susSYK}) when $U_p=u+v\cos(p)$. Now we check (\ref{largeNsuppress}) in SYK chain. By using the effective action in Appendix A, we can write the left-hand side of (\ref{largeNsuppress}) as
\bea\label{S0BO}
    \int \mathcal{D}G\mathcal{D}\Sigma \, e^{-S_0[G,\Sigma]+ \sum_s\iint d\tau_1 d\tau_2 \frac{C^2}{4}  B_s(\tau_1)  G^s(\tau_1,\tau_2)^2 B_s(\tau_2)},
\eea
where $S_0[G,\Sigma]$ is the undeformed effective action for bi-local field $G^s(\tau_1,\tau_2)$ and $\Sigma^s(\tau_1,\tau_2)$, which is order $N$, while the second term $\sum_s\iint d\tau_1 d\tau_2 \frac{C^2}{4}  B_s(\tau_1)  G^s(\tau_1,\tau_2)^2 B_s(\tau_2)$ is order $1$. Thus, at leading order of $1/N$ expansion, we can just replace the bi-local field in (\ref{S0BO}) by its saddle point solutions, which are separately evaluated by using $S_0[G,\Sigma]$. This directly leads to right-hand side of (\ref{largeNsuppress}) where $\mathcal{G}^s_0(\tau)=\frac{C^2}2 G_c(\tau)^2$.

\subsection{Double trace deformation from boundary condition in AdS$_2$ chain}
The Minkovski action in the background $ds^2= \frac{1}{z^2}(-dt^2+dz^2)$ is
\bea
	S &=& -\frac12 \sum_s \int d^2 x \sqrt{-g} \Big( \partial_\mu \chi_{s} \partial^\mu \chi_s + m^2 \chi_{s} \chi_s \Big)+ \frac{2\nu-1}4  \sum_s \int_{z=\epsilon} dt \sqrt{-h}\chi_s^2 \\
	&& + \frac12\zeta^2\epsilon^{2\nu}\sum_s \int_{z=\epsilon} dt\sqrt{-h} (u\chi_s^2+v\chi_s\chi_{s+1}) + \zeta\epsilon^{\frac12+\nu}\sum_s\int_{z=\epsilon} dt\sqrt{-h} J_s\chi_s.
\eea
We perform a Fourier transformation $\chi_s= \frac{1}{\sqrt{M}} \sum_p \chi_p e^{ips}$, where $M$ is the number of sites. The EOM with boundary conditions are
\bea
	&& \frac{1}{\sqrt{-g}} \partial_\mu (\sqrt{-g} g^{\mu\nu} \partial_\nu \chi_p)- m^2\chi_p = 0, \\
	&& n \cdot \partial \chi_p+  (\frac12- \nu - \zeta^2 \epsilon^{2\nu}U_p) \chi_p= \zeta \epsilon^{\nu+\frac12} J_p,
\eea
where $n=n^z=-\epsilon$ and $U_p\equiv u+v \cos(p)$. Thus, the on-shell action is $S = \frac12 \sum_p \int dx \sqrt{-h} \zeta \epsilon^{\frac12+ \nu} \chi_p J_{-p}$.

A general solution of EOM with infalling boundary condition near boundary $z\rightarrow \infty$ is given by $ \chi_p= A_p z^{\frac12-\nu}+ B_p z^{\frac12+\nu} $ with $\eta(\omega)=\frac{A_p}{B_p}=2^{2\nu} \frac{\Gamma(\nu)}{\Gamma(-\nu)}(-i\omega)^{-2\nu}$.  And the boundary condition can be simplified as $ -2\nu B_p- \zeta^2 U_p A_p=\zeta J_p $. Thus we get $A_p= -\frac{\zeta}{\zeta^2 U_p + \frac{2\nu}{\eta(\omega)}}J_p $. The generating functional is
\bea
	S &=& -\frac12 \sum_p \int dx J_{-p}\frac{\zeta^2}{\zeta^2 U_p + \frac{2\nu}{\eta(\omega)}}J_p.
\eea
The propagator is given by $ \mathcal{G}_{u,v}(\omega, p) = -\frac{\zeta^2}{\zeta^2 U_p+ \frac{2\nu}{\eta(\omega)}}=\frac{1}{-U_p+ \mathcal{G}_{0,0}(\omega,p)^{-1}} $.

Finite temperature corresponds to AdS$_2$ black hole background, i.e., $ds^2= \frac{L^2}{z^2}\Big(-(1-\frac{z^2}{z_0^2})dt^2+ \frac{1}{1-\frac{z^2}{z_0^2}} dz^2 \Big) $. The horizon is at $z=z_0$, and temperature is given by $T= \frac{1}{2\pi z_0}$. Following the same steps, we can get
\bea
	\eta(\omega, z_0)=(z_0/2)^{2\nu} \frac{\Gamma(2\nu) \Gamma(\frac12- \nu- i \omega z_0) \Gamma(\frac12- \nu)}{\Gamma(-2\nu) \Gamma(\frac12+ \nu- i \omega z_0) \Gamma(\frac12+ \nu)}.
\eea
for finite temperature, which leads to the propagator shown in the main text.

\subsection{Marginal case: $q=4$}
In the SYK$_4$ chain, with the background solution given by $ G^s_f(\tau_1,\tau_2)= b \frac{\text{sgn}(\tau_1-\tau_2)}{\sqrt{|\tau_1-\tau_2|}}$, the action involving order parameter becomes
\bea
	S &=& \sum_p \int \Big( O_{-p} (-\frac{u}{2} -\frac{v}{2} \cos p )O_p+ B_{-p} O_p  +  \frac{b^2C^2}{2} B_{-p} (-\omega)  \ln|\omega| B_p(\omega) \Big).
\eea
Integrating out $B$ fields to get $ S = \frac12 \sum_p \int O(-\omega) \Big (-u-v\cos p- \frac{1}{b^2 C^2} \frac{1}{\ln |\omega|} \Big) O(\omega) $, one get the dynamic susceptibility,
\bea
	\mathcal{G}(i \omega,p)= \frac{1}{-u-v\cos p- \frac{1}{b^2 C^2} \frac{1}{\ln |\omega|}} =\frac{1}{-u-v\cos p- \frac{\sqrt{4\pi J^2}}{C^2} \frac{1}{\ln |\omega|}}.
\eea

In the holographic calculations, without double-trace deformation, $U_p=0$, we have
\bea
	\mathcal{G}^R_{0,0}(\omega, p)= \frac{\zeta^2}{4} \times 2^{2\nu+1} \frac{\Gamma(\nu)}{-\nu\Gamma(-\nu)}(-i\omega)^{-2\nu}.
\eea
Since we cannot directly take $\nu$ to zero, we transform it back to real space,
\bea
	\mathcal{G}^R_{0,0}(t,p)= \frac{\zeta^2}{4} \times\frac{2}{\pi^{d/2}} \frac{\Gamma(\frac{d}{2}-\nu)}{\Gamma(1-\nu)} \frac{1}{|t|^{d-2\nu}},
\eea
which is consistent with Witten's refined formula \cite{Klebanov:1999tb}. Then this formula can be continued to $\nu=0$ without any zero, i.e.,
\bea
	\mathcal{G}^R_{0,0}(t,p)= \frac{\zeta^2}{4} \frac{2}{\pi^{1/2}} \frac{\Gamma(\frac{1}{2})}{\Gamma(1)} \frac{1}{|t|}.
\eea
giving rise to $ \mathcal{G}_{0,0}(i \omega)=-\zeta^2 \ln |\omega|$. After turning on the deformation, we get
\bea\label{marginalG}
\mathcal{G}_{u,v}(i\omega, p) =\frac{1}{-u-v \cos p+ \mathcal{G}_{0,0}(i\omega,p)^{-1}}= \frac{1}{-u-v\cos p- \frac{\sqrt{4\pi J^2}}{C^2} \frac{1}{\ln |\omega|}},
\eea
which is identical to the dynamic susceptibility in SYK$_4$ chain. The instability analysis is the same as the main text, except that now the stable fixed point and the Gaussian fixed point coincide and the double-trace deformation becomes marginally irrelevant(relevant) for $u<-|v|$($u>-|v|$) once we take the $q=4$ (or equivalently $\nu=0$) in the RG equations in the main text. So the QCP is called marginal QCP in Ref. \cite{Iqbal:2011aj}.

We briefly introduce the calculation in the bulk in marginal case. Notice that when $q=4$, there are $\nu=0$ and $m^2L^2=-\frac14$. The mass of the scalar fields saturates the Breitenlohner-Freedman bound of AdS$_2$. Near the boundary, $\chi_p$ behaves as $\chi_p \to A_p z^{\frac12}\ln(z)+B_p z^{\frac12}$. The terms in (\ref{actionscalar}) should be modified to counter the logarithmic divergence in $S_\chi$ \cite{Skenderis:2002wp} and give a well defined boundary condition for variational problem. They can be replaced by
\bea
  S_{ct} &=&\left(-\frac14 - \frac12 \frac1{\ln(\epsilon)}\right)  \sum_s \int_{z=\epsilon} dt \sqrt{-h}\chi_s^2, \\
  S_{u,v} &=& \frac12\zeta^2\frac1{\ln(\epsilon)^2}\sum_s \int_{z=\epsilon} dt\sqrt{-h} (u\chi_s^2+v\chi_s\chi_{s+1}), \\
  S_J &=& \zeta\frac{\sqrt{\epsilon}}{\ln(\epsilon)}\sum_s\int_{z=\epsilon} dt\sqrt{-h} J_s\chi_s.
\eea
Following the similar derivation in Appendix E, one will obtain the Green's function which coincides with (\ref{marginalG}).

\subsection{Double-trace deformation in Reissner-Nordstrom AdS black hole}

We consider a four-dimensional Reissner-Nordstrom (RN) AdS black hole with metric
\bea
	ds^2=- f(r) dt^2+ \frac{1}{f(r)} dr^2+ r^2 d\Omega^2, \quad f(r)=1+\frac{Q^2}{r^2}- \frac{Q^2}{r r_+}- \frac{r_+^3}{r L^2}+ \frac{r^2}{L^2},
\eea
where $L$ is the AdS radius, $Q$ is electric charge, and $r_+$ is the horizon. The temperature is given by $T= \frac{1+\frac{3r_+^2}{L^2}- \frac{Q^2}{r_+^2}}{4\pi r_+} $.

Near horizon geometry is AdS$_2 \times S^2$ with AdS$_2$ radius given by $L_2$. In the following we consider the limit $Q\gg L$, then $L_2^2 \approx \frac{L^2}{6}$. Turning slightly away from extremity, $r_+ \approx 3^{-1/4} (LQ)^{1/2}+ \frac{\pi}{3} L^2 T \epsilon$, and using coordinate $r=r_+ + \epsilon \hat r$, $t= \hat{t}/\epsilon$, we expand the metric with respect to $\epsilon$ to get
\bea
	ds^2 = -\frac{\hat{r}^2-(\frac{\pi}{3} L^2 T)^2}{L_2^2} d\hat t^2+ \frac{L_2^2}{\hat{r}^2-(\frac{\pi}{3} L^2 T)^2} d\hat r^2+ r_+^2 d\Omega^2.
\eea
To connect to the coordinate used in calculations before, we make further coordinate transformation $z= \frac{L^2_2}{\hat r}$, the metric becomes
\bea\label{metric}
ds^2=  \frac{L^2_2}{z^2}\Big[ -(1-\frac{z^2}{z_0^2})dt^2+ \frac{1}{1-\frac{z^2}{z_0^2}} dz^2 \Big]+ r_+^2 d\Omega^2,
\eea
where $z_0= \frac1{2\pi T}$ and the metric in $t, z$ component is same as the AdS$_2$ black hole used in the calculation of Green's function.

Now we consider a free scalar with mass $m$ in this RN AdS black hole background,
\bea
	I= \int d^4 x \sqrt{-g} \Big( -\frac12 \partial_\mu \chi \partial^\mu \chi- \frac12 m^2 \chi^2 \Big),
\eea
Assume $\chi= \chi(r) Y_{lm'}(\theta,\phi) e^{-i\omega t}$, where $Y$ is the spherical harmonics, the EOM can be reduced to
\bea
	\partial_r (r^2 f(r) \partial_r \chi)+ \Big( \frac{r^2\omega^2}{f(r)} -l(l+1)- r^2 m^2 \Big)\chi=0.
\eea
Below, we focus on spherical scalar field, i.e., $m'=l=0$, for simplicity. According to mass-scaling dimension relation, we have the asymptotic behaviors $ \chi \sim \alpha r^{-\Delta_-}+ \beta r^{-\Delta_+}, \chi \sim \hat{\alpha} (r-r_+)^{-\delta_-}+ \hat{\beta} (r-r_+)^{-\delta_+}$, where $\Delta_\pm= \frac32 \pm \sqrt{\frac94+m^2 L^2}$ and $\delta_\pm= \frac12 \pm \sqrt{\frac14+m^2 L_2^2}$. At the limit $T, \omega \rightarrow 0$, we denote $\alpha, \beta \rightarrow \alpha_0, \beta_0$, $\hat \alpha, \hat \beta \rightarrow \hat \alpha_0, \hat \beta_0$. There is a relation between those coefficients
\bea
	\left( \ba{cccc}  \alpha_0 \\ \beta_0 \ea \right) = M \left( \ba{cccc}  \hat \alpha_0 \\ \hat \beta_0 \ea \right),
\eea
where $M= \left( \ba{cccc} a^+ & a^- \\ b^+ & b^- \ea \right) $, and we define $u_c= \frac{b^+}{a^+}$.

We separate the calculations into inner and outer region. In the inner region, we have background metric given by (\ref{metric}) and the solution is given by $\mathcal{G}_R(\omega, T)= \frac{\hat \alpha}{\hat \beta}$. In the outer region, let $\omega \rightarrow \epsilon \omega $, $T \rightarrow \epsilon T$, we can expand the solution in terms of $\epsilon$, $\chi=\chi_0+ \epsilon \chi_1+ ...$, $f=f_0+ \epsilon f_1+ ...$. Then (note that we only keep $T$ to linear order)
\bea
	D \chi_0=0, \quad D\chi_1=- \partial_r r^2 f_1 \partial_r \chi_0, \quad D\chi_2= -  \frac{r^2 \omega^2}{f_0} \chi_0+ O(T^2),
\eea
where $D= \partial_r r^2 f_0 \partial_r -r^2 m^2$ and $ f_0(r)= \frac{Q^2}{r^2}+ \frac{r^2}{L^2}- \frac{4Q^{3/2}L^{-1/2}}{3^{3/4}r}, f_1(r) \approx - \frac{\pi L^2 T}{3 r}$.
Let $\chi_L= \chi_1+\chi_2$ as correction to the zeroth-order scalar, then it satisfies
\bea
	D \chi_L= \frac{\pi L^2 T}{3} \partial_r (r \partial_r \chi_0) -  \frac{r^2 \omega^2}{f_0} \chi_0.
\eea
We can manipulate such that
\bea
	\chi_L \sim A_1 (r-r_+)^{-\delta_--1}+ A_2 (r-r_+)^{-\delta_--2}+ ... \\
	B_1 (r-r_+)^{-\delta_+-2} + B_2 (r-r_+)^{-\delta_+-2}+ ....
\eea
Then it is obvious to have $ \chi_0 =\hat{\alpha}_0 (r-r_+)^{-\delta_-}+ \hat{\beta}_0 (r-r_+)^{-\delta_+}$.

The Green's function under double-trace deformation is given by
\bea
	G_R= \frac{\alpha}{-u \alpha + \beta}=\frac{\alpha_0 + \alpha_L}{-u \alpha_0 + \beta_0-u \alpha_L + \beta_L}.
\eea
We know that
\bea
	-u \alpha_0 + \beta_0 \approx \hat \alpha_0( -a^+ (u-u_c) + \frac{\det M}{a^+} \mathcal{G}_R^{-1} ).
\eea
To get $-u \alpha_L + \beta_L$, we consider $\chi_{u_c}$ that satisfies
\bea
	D\chi_{u_c}=0, \quad \chi_{u_c} \sim (r-r_+)^{-\delta_-}, \quad \chi_{u_c} \sim a^+(r^{-\Delta_-}+ u_c r^{-\Delta_+}),
\eea
then
\bea
	\int dr \chi_{u_c} D \chi_L=[\chi_{u_c} r^2 f_0 \chi_L'- \chi_{u_c}' r^2 f_0 \chi_L]_{r_+}^\infty+ \int dr \chi_L D \chi_{u_c}= - \frac{a^+(-u \alpha_L + \beta_L)(\Delta_+ - \Delta_-)}{L^2}.
\eea
Thus we have
\bea
- \frac{a^+(-u \alpha_L + \beta_L)(\Delta_+ - \Delta_-)}{L^2}= \frac{\pi L^2 T}{3} \int dr \chi_{u_c} \partial_r (r \partial_r \chi_0) - \omega^2  \int dr \chi_{u_c} \frac{r^2 }{f_0} \chi_0 \approx \hat \alpha_0 ( a_T T- a_\omega \omega^2  ),
\eea
where $a_T=  \frac{\pi L^2 }{3} \int dr \chi_{u_c} \partial_r (r \partial_r \chi_{u_c})$ and $a_\omega= \int dr  \frac{r^2 }{f_0} \chi_{u_c}^2 $. Combining all the results, we have
\bea
	G_R(\omega, T)= \frac{Z}{-u_\text{IR}+ \mathcal{G}_R^{-1}(\omega, T)+ c_\omega \omega^2- c_T T},
\eea
where $u_\text{IR}= \frac{(a^+)^2}{\det M}(u-u_{c}) $, $Z= \frac{(a^+)^2}{\det M}$, $c_\omega= \frac{(a^+)^2L^2}{\det M(\Delta_+- \Delta_-)} a_\omega $, and $c_T= \frac{(a^+)^2L^2}{\det M(\Delta_+- \Delta_-)} a_T $. In the transitions of the single-site SYK model, we know that $\nu= \frac12 - \frac4q$. As a result, $c_\omega, c_T$ and higher-order corrections are irrelevant in low energy limit, i.e., $G_R(\omega, T) \approx \frac{Z}{-u_\text{IR}+ \mathcal{G}_R^{-1}(\omega, T)}$, which resembles the susceptibility of the transition in the SYK model. Actually, the corrections from perturbative calculations in the outer region is analytical in many higher-dimensional black hole. Thus, the transitions in SYK model captures the universal properties of double-trace deformations in generic black holes with near horizon AdS$_2$ geometry.

\subsection{Semi-holographic effective field theory}
The transition induced by four-fermion interactions in the SYK chain can also be understood by effective field theory of semiholographic type \cite{Faulkner:2010tq,Jensen:2011af} (let's focus on $v>0$ case; $v<0$ is similar). The effective Lagrangian consists three parts: a Landau-Ginzburg (LG) theory of the order parameter $\phi(\tau,x)$ describing the $Z_2$ symmetry breaking, an emergent conformal sector resulted from the strongly interacting Majorana fermions degrees of freedom, i.e., $\mathcal{L}_\text{IR}[\Psi]$, where $\Psi$ refers to the emergent large-$N$ degrees of freedom, and a coupling part between above two degrees of freedom,
\bea
 \mathcal{L}[\phi, \Psi] &=& \mathcal{L}_\text{LG}[\phi]+ \mathcal{L}_\text{IR}[\Psi]+\mathcal{L}_c[\phi,\Psi].
\eea
The LG theory contains all symmetry-allowed terms, but one can restrain on the first few orders, i.e.,
\bea
	\mathcal{L}_\text{LG}[\phi] &=& \frac12 c_t^2 (\partial_\tau \phi)^2+ \frac12 c_x^2 (\partial_x \phi)^2+ \frac r2 \phi^2,
\eea
where $r$ is the tuning parameter, i.e., $r=-u-v$ and $c_t, c_x, u$ are constants resulted from integrating out some UV data. Actually, the conventional $Z_2$ transition in two-dimension is described by Ising conformal field theory. It can be inferred from the fact that $\phi$ has vanishing scaling dimension and one should actually retain infinite terms. However, the $Z_2$ symmetry-breaking transition in the SYK chain falls into a different universality class, and one can see that the above truncation correctly captures the universal features here. Note that the continuous $x$ coordinate is obtained by a continuum approximation of the site index $s$ in the SYK chain.

The conformal sector is defined through holography by
\bea
	\mathcal{L}_\text{IR}[J_\Psi] &=& \lim_{\epsilon \rightarrow 0} e^{-S_\text{grav}[\psi^{cl}]}|_{\psi^{cl}(\epsilon,\tau,x)= \epsilon^{\Delta_-} J_\Psi(\tau,x)},
\eea
where $J_\Psi$ is the source of field $\Psi$ and $\psi$ is the bulk field dual to $\Psi$ and is subjected to standard quantization. In our case, the gravity action is defined as the matter sectors on the AdS$_2$ chain,
\bea
	S_\text{grav}[\psi]&=&\sum_s \frac12\int dx^2\sqrt{g_s} (g_s^{\mu\nu}\partial_\mu\psi_s\partial_\nu\psi_s + m^2 \psi_s^2 ).
\eea
Using the uniform AdS$_2$ chain metric, i.e., $ds_s^2=\frac{1}{z^2}(d\tau^2+ dx^2)$, the conformal sector gives rise to the propagator of $\Psi$ field without couplings to the order parameter,
\bea
	G_\Psi^{(0)}(\omega, k)= c_\Delta |\omega|^{2\Delta-1},
\eea
where $c_\Delta\equiv  -\frac{2(\Delta-\frac12)}{2^{2(\Delta-\frac12)}} \frac{\Gamma(\frac32-\Delta)}{\Gamma(\frac12+\Delta)}$ and $\Delta= \frac12+ \sqrt{\frac14+ m^2}$. One can see that the correlator is independent of momentum $k$, which is one of the essential features of the AdS$_2$ chain.

Finally, the couplings between these two sectors are given by
\bea
	\mathcal{L}_c[\phi, \Psi] &=& \zeta \phi \Psi,
\eea
where $\zeta$ is a constant. From the coupling part one can infer that $\phi$ appears like the source of $\Psi$. So once we integrate out the $\Psi$ field, to the first order of $1/N$ the effective action for the order parameter is given by
\bea
	\mathcal{L}[\phi]= \mathcal{L}_{GL}[\phi]+ \frac{\zeta^2}2 \phi G_\Psi^{(0)}\phi,
\eea
which gives the susceptibility
\bea
	\mathcal{G}(\omega, k)= \frac{1}{c_t^2 \omega^2+ c_x^2 k^2 + r + \zeta^2 c(\Delta) |\omega|^{2\Delta-1} }.
\eea
which matches exactly the susceptibility from calculations in $\Delta=1-\frac2q$ and $
\zeta^2=\frac{\sqrt{\pi } C^2 b^2 \Gamma \left(\nu +\frac{1}{2}\right)}{\tan (\pi  \nu )\Gamma (\nu )}$. For a general scaling dimension $\Delta$, the critical exponents are shown in Table A1. One can see that they reproduce the critical exponents in the SYK chain for $\Delta=1-\frac2q$ as expected. The IR conformal sector in the semi-holography theory originates from the strongly interacting Majorana fermions. The emergent large-$N$ field $\Psi$ can be identified to the field $B$ in the SYK model.

There are a few comments contrasting the full-holographic theory and semi-holographic theory. We choose to use full holographic theory because the order parameter $O$ behaves like a large-$N$ single-trace operator in the SYK chain, and is naturally dual to the bulk field $\chi$ defined in the manuscript. The full-holographic construction is straightforward.

On the other hand, in the semi-holographic theory, the order parameter, i.e., the $\phi$ field, only lives in the boundary and does not have a bulk dual. As a result, the LG theory in the semi-holographic model is uniquely determined by symmetry, without the knowledge about the large-$N$ structure. The neglect of the higher-order terms (higher than quadratic order) in the LG theory is not a result of fine tuning, but a result of the suppression by $1/N$. The knowledge that higher order terms (higher than quadratic order) in the LG theory are suppressed by $1/N$ factor cannot be obtained within the semi-holographic framework and should be put in by hand via matching from the results of the SYK chain. This also implies that the critical exponents would receive corrections once we include the effects to the next order in $1/N$.

In the large-$N$ limit and in the two-point function level, both the full-holographic theory represented in the manuscript and the semi-holographic theory represented here can reproduce exactly the same susceptibility in the SYK chain. Ultimately, both full-holographic theory and semi-holographic theory should work for the QCP in the SYK chain.
\end{widetext}

\end{document}